\documentclass[%
reprint,
showpacs,preprintnumbers,
 amsmath,amssymb,
 aps,
prb,
]{revtex4-1}

\usepackage{graphicx}% Include figure files
\usepackage{dcolumn}% Align table columns on decimal point
\usepackage{bm}% bold math
\usepackage{color}
\newcommand*{\rttensor}[1]{\overline{\overline{#1}}}
\newcommand*{\rtvector}[1]{\overline{#1}}
\begin{document}

\preprint{APS/123-QED}

\title{Quantification of uncertainty in first-principles predicted mechanical properties of solids: Application to solid ion conductors}% Force line breaks with \\

\author{Zeeshan Ahmad}
\author{Venkatasubramanian Viswanathan}%
 \email{venkvis@cmu.edu}
\affiliation{%
 Department of Mechanical Engineering, Carnegie Mellon University, Pittsburgh, Pennsylvania 15213\\
}

\date{\today}

\begin{abstract}
Computationally-guided material discovery is being increasingly employed using a descriptor-based screening through the calculation of a few properties of interest.  A precise understanding of the uncertainty associated with first principles density functional theory calculated property values is important for the success of descriptor-based screening.    Bayesian error estimation approach has been built-in to several recently developed exchange-correlation functionals, which allows an estimate of the uncertainty associated with properties related to the ground state energy, for e.g. adsorption energies.  Here, we propose a robust and computationally efficient method for quantifying uncertainty in mechanical properties, which depends on the derivatives of the energy.  The procedure involves calculating the energy around the equilibrium cell volume with different strains and fitting the obtained energies to the corresponding energy-strain relationship. At each strain, we use instead of a single energy, an ensemble of energies, giving us an ensemble of fits and thereby, an ensemble of mechanical properties associated with each fit, whose spread can be used to quantify its uncertainty. The generation of ensemble of energies is only a post-processing step involving a perturbation of parameters of the exchange-correlation functional and solving for the energy non-self consistently. The proposed method is  computationally efficient and provides a more robust uncertainty estimate compared to the approach of self-consistent calculations employing several different exchange-correlation functionals.  We demonstrate the method by calculating the uncertainty bounds for Si using the developed method.  We show that the calculated uncertainty bounds the property values obtained using three different GGA functionals: PBE, PBEsol and RPBE.  Finally, we apply the approach to calculate the uncertainty associated with the DFT-calculated elastic properties for solid state Li-ion and Na-ion conductors.

\end{abstract}

\pacs{62.20.-x, 71.15.Nc, 46.25.-y}% PACS, the Physics and Astronomy
\maketitle

%\tableofcontents

\section{\label{sec:level1}Introduction}
Material innovation is at the heart of developing new tools and technologies to address the societal needs for clean energy and human health.\cite{holdren2014materials}  First-principles density functional theory (DFT) calculations have played a crucial role in accelerating material innovation by allowing the prediction of chemical, \cite{KohnDFTReview, neese2009prediction} mechanical, \cite{ ElasticDFT, ZeoliteMech, Hasnip20130270} and electrical \cite{Muscat2001397, Hasnip20130270} properties of materials. DFT calculations have been employed to identify new battery electrodes,\cite{Kang977} photovoltaics,\cite{AlanPhoto} catalysts\cite{Nørskov18012011}, thermoelectrics\cite{tang2015convergence} etc. An approach to computationally-guided material discovery is to employ a descriptor-based search where materials are screened for a few properties like band gap, \cite{C1EE02717D} adsorption energy,\cite{ClimbAcVol} HOMO levels, \cite{khetanHOMO} and the identified candidates are synthesized, characterized and tested for their functionality.  Given the time and resource consumed for experimental testing and validation, there is a growing realization that it is crucial to quantify the uncertainty associated with the DFT-predicted property values.

The reliability of DFT calculations is typically estimated through comparisons to experiments or to data sets of higher-level calculations. \cite{ErrorDFTExp, benchmarkCC, CompDFTCCT} 
%Comparison of DFT-predicted values for, for example, cohesive energy and elastic properties of crystalline solids with experiments revealed a systematic trend that DFT predicts certain classes of elements better than others. \cite{ErrorDFTExp} However, the study was limited to GGA-PBE exchange-correlation functional, \cite{GGAPBE} while hundreds of GGA exchange-correlation functionals are being used for DFT calculations. 
Studies using different exchange-correlation functionals have shown considerable variation in properties like adsorption energy, \cite{AdsorpEnergy} crystal structure of water, \cite{CrystalWater} elastic constants, \cite{Deng01012016} vibrational frequencies, \cite{VibFreq} thermal conductivity \cite{ThermalCond}, infrared spectrum \cite{IRSpec} etc. It is therefore of great interest to isolate the error associated with the exchange-correlation functional when comparing DFT-predicted values with experimental values. A recently developed exchange-correlation functional, Bayesian error estimation functional with van der Waals correlation (BEEF-vdW) possesses built-in error estimation capabilities. \cite{wellendorff2012density}  The Bayesian error estimation \cite{BEE2005} within the functional is designed to reproduce known energetic errors by mapping the uncertainties on the exchange-correlation parameters.  This capability has been exploited to estimate the uncertainty in adsorption energies and thereby the reliability of calculated catalytic rates for ammonia synthesis \cite{Medford197} and electrocatalytic oxygen reduction. \cite{Deshpande2016}

The calculation of uncertainties within the Bayesian error estimation approach has been limited to quantities that are directly related to the ground-state energy. \cite{Wang2016, RuneHeine, li2016activating, Pandey2015} In this work, we propose a method to calculate the uncertainty in properties that involve the derivatives of energy, for e.g., mechanical properties.  Specifically, we demonstrate the method to estimate the uncertainty associated with the calculated elastic properties for solids. This is done by performing an ensemble of energy-strain fits around equilibrium. The elastic constants for each fit can be calculated in terms of the fitting parameters, and the spread of their distribution can be used to quantify the uncertainty associated with the elastic constants.

We use the developed method to calculate the mechanical properties with uncertainty for candidate solid ion conductors for Li and Na-ion batteries.  It has been shown that solid ion conductors that possess a sufficient modulus can suppress the formation of dendrites at the metal anode. \cite{Monroe01022005}  Here, we focus on four important classes of solid ion conductors: thiophosphate, halide, antiperovskite and glass. We use the ensemble of obtained mechanical property values to determine other properties of interest like the Pugh's modulus ratio with uncertainty bounds.

\section{Methods}

\subsection{Property Calculation}
\label{sec:propcalc}
The elastic constants of a material can be obtained by computing a set of energies for its unit cell at different strains using DFT calculations.   The calculated energies can be fit to the energy-strain relationship and the elastic constants can be extracted from the fitting parameters. Several choices exist regarding the strains to be applied. We outline the procedure used in this section.

We assume the undeformed coordinates of a point in the material as $ \rtvector{X}=(X_1,X_2,X_3)^T$, where $T$ denotes the transpose. The coordinates are transformed on applying a homogeneous deformation $\rttensor{F}$ such that the new coordinates $\rtvector{x}=(x_1,x_2,x_3)^T$ are given by $\rtvector{x}=\rttensor{F} \rtvector{X}$. From the deformation matrix $\rtvector{F}$, we get the Lagrangian strain tensor, $\rttensor{\eta}$, given by
\begin{equation}
\rttensor{\eta}=\frac{1}{2}( \rttensor{F}^T \rttensor{F} - \rttensor{I} ).
\end{equation}

The energy $E$ of the unit cell having a volume $V$ on applying a Lagrangian strain $\rttensor{\eta}$ can be expressed in terms of the elastic constants $C_{ijkl}$ as
\begin{equation} \label{eq:energyexp}
E(\rttensor{\eta})=E_0+\frac{V}{2}\sum_{ijkl}C_{ijkl}\eta_{ij}\eta_{kl}+O(\eta_{ij}^3),
\end{equation}
where $E_0$ denotes the energy at equilibrium or zero strain. Since all energy calculations using DFT are performed at 0 K, all elastic constants in Eq. (\ref{eq:energyexp}) are isothermal constants at 0 K:
\begin{equation}
C_{ijkl}(\text{T=0 K})=\frac{1}{V} \left. \frac{\partial^2 E}{\partial \eta_{ij} \partial \eta_{kl}}\right|_{\rttensor{\eta}=0}.
\end{equation}

Using the Voigt notation for indices \cite{wallace1998thermodynamics} ($11\rightarrow1$, $22\rightarrow2$, $33\rightarrow3$, $32$ or $23\rightarrow4$, $13$ or $31\rightarrow5$, $12$ or $21\rightarrow6$), the 4th order elastic tensor can be written in a contracted form as a 2nd order $6\times6$ tensor. The deformation matrix we choose is of the form,  $\rttensor{F}=\rttensor{I}+\rttensor{\epsilon}$, where $\rttensor{\epsilon}$ is a symmetric matrix with six independent components: \cite{mehl1994first}
\begin{equation}
\rttensor{\epsilon}=\begin{bmatrix}
e_1 & e_6/2 & e_5/2\\
e_6/2 & e_2 & e_4/2\\
e_ 5/2& e_4/2 & e_3
\end{bmatrix}.
\end{equation}

For crystals having cubic symmetry, the elastic tensor has only three independent components: $C_{11}$, $C_{12}$ and $C_{44}$, and only three independent strains are required. On applying a volume-conserving orthorhombic strain \cite{mehl1994first}
\begin{equation}
e_1 = -e_2 = x, \\
e_3 = \frac{x^2}{1-x^2},\\
e_4=e_5=e_6=0,
\end{equation}
the energy expansion calculated using Eq. (\ref{eq:energyexp}) is
\begin{equation}
E(x)=E_0+V(C_{11}-C_{12})x^2+O(x^4),
\end{equation}
which can be used to obtain the elastic constant $C_{11}-C_{12}$. Similarly, the energy change due to a volume-conserving monoclinic strain \cite{mehl1994first}
\begin{equation}
e_6 =  x, \\
e_3 = \frac{x^2}{4-x^2},\\
e_1=e_2=e_4=e_5=0,
\end{equation}
can be calculated as
\begin{equation}
E(x)=E_0+\frac{V}{2}C_{44}x^2+O(x^4).
\end{equation}
This gives the elastic constant $C_{44}$. Further, a uniform strain in all three directions can be used to calculate the bulk modulus $B=(C_{11}+2C_{12})/3$ and lattice constant by fitting the energies to the Birch-Murnaghan equation of state. \cite{BMEOS1947}

For lower symmetry crystals, more strain-energy calculations are required. \cite{le2001symmetry, le2002symmetry, mehl1994first, zhao2007first} Since the only lower symmetry crystal in our calculations has tetragonal symmetry, we show the strains used for this case in Table \ref{tab:tetra}.

\begin{table}[b]
\caption{\label{tab:tetra}%
Strains used determine the elastic constants of crystals with tetragonal symmetry}
\begin{ruledtabular}
\begin{tabular}{lcc}
\textrm{Strain}&
\textrm{Non-zero $e_i$}&
\textrm{Energy change $E(x)-E_0$}\\
\colrule
1 & $e_1=e_2=x$ & $V(C_{11}+C_{12})x^2+O(x^3)$\\
2 & $e_1=e_2=x$, &  $(C_{11}+C_{12}+2C_{13}-4C_{33})x^2$\\ 
& $e_3=\frac{-x(2+x)}{(1+x)^2}$ & $+O(x^3)$\\
3 & $e_3=x$ & $C_{33}x^2/2+O(x^3)$\\
4 & $e_1=[(1+x)/(1-x)]^{1/2}-1$, & $(C_{11}-C_{12})x^2+O(x^4)$ \\
& $e_2=[(1-x)/(1+x)]^{1/2}-1$\\
5 & $e_4=e_5=x, e_3=x^2/4$ & $C_{44}x^2+O(x^4)$\\
6 & $e_6=x$, & $C_{66}x^2/2+O(x^4)$\\
& $e_1=e_2=(1+x^2/4)^{1/2}-1$ 
\end{tabular}
\end{ruledtabular}
\end{table}

We used the Voigt-Reuss-Hill approximation \cite{VRHApprox} to relate the polycrystalline bulk, shear and Young's modulus to the single crystal elastic constants. The Voigt and Reuss approximations for bulk modulus of a cubic polycrystal are the same: $B=(C_{11}+2C_{12})/2$ whereas the shear modulus is given by \cite{VRHapproxcubic}
\begin{subequations}
\begin{eqnarray}
G_{V}=\frac{C_{11}-C_{12}}{5}+\frac{3C_{44}}{5},\\
G_R=\frac{5(C_{11}-C_{12})}{4C_{44}+3(C_{11}-C_{12})}.
\end{eqnarray}
\end{subequations}
The bulk and shear modulus of a tetragonal crystal under the Voigt and Reuss approximations can be calculated using \cite{SisodiaPoly}
\begin{subequations}
\begin{eqnarray}
B_V=\frac{1}{9}(2C_{11}+C_{33}+2C_{12}+4C_{13}),\\
B_R=\frac{(C_{11}+C_{12})C_{33}-2C_{13}^2}{C_{11}+C_{12}+2C_{33}-4C_{13}}.
\end{eqnarray}
\end{subequations}
\begin{subequations}
\begin{gather}
\begin{split}
G_V &=\frac{1}{30}\left[4(C_{11}-C_{13})+2(C_{33}-C_{12})\right.\\
&\left.+6(C_{66}+2C_{44})\right],\\
\end{split}\\
\begin{split}
G_R=&\frac{30}{4} \left[\frac{2(C_{11}+C_{12})+C_{33}+4C_{13}}{(C_{11}+C_{12})C_{33}-2C_{13}^2}\right.\\
&\left. +\frac{3}{C_{11}-C_{12}}+\frac{3}{2C_{66}} +\frac{3}{C_{44}} \right]^{-1}.
\end{split}
\end{gather}
\end{subequations}
The average of the Voigt and Reuss limits was used as the polycrystalline bulk and shear modulus.
\begin{subequations}
\begin{eqnarray}
B = \frac{B_V+B_R}{2},\\
G=\frac{G_V+G_R}{2}.
\end{eqnarray}
\end{subequations}
The Young's modulus $Y$ and Poisson's ratio $\nu$ can be calculated from the bulk and shear modulus using
\begin{eqnarray}
Y=\frac{9BG}{3B+G},\\
\nu=\frac{3B-2G}{2(3B+G)}.
\end{eqnarray}
\subsection{Computational Details}

Self consistent density functional calculations were performed in the real-space projector-augmented wave (PAW) method \cite{BlochlPAW1994, KresseJoubert1999} as implemented in \texttt{GPAW}. \cite{mortensen2005real, enkovaara2010electronic} We employed the BEEF-vdW exchange-correlation functional \cite{wellendorff2012density} with 2000 ensembles for all calculations. A real space grid spacing of 0.14 $\mathrm{\AA}$ and Monkhorst Pack\cite{Monkhorstkpoint} scheme for sampling the Brillouin zone was used. All calculations were converged to energy $< 0.1$ meV for the unit cell and force $< 0.01$ eV/$\mathrm{\AA}$. The k-point density was optimized for individual structures to achieve the desired energy convergence. The strain parameter $x$ was varied between -5 to 5\% and all energy-strain fittings were performed such that the fitting parameters were converged with respect to the maximum value of $x$ used in the fit. The degree of the polynomial used for fitting was three or four depending on the energy-strain relationship.

\subsection{Bayesian Error Estimation}
The Bayesian error estimation functional with van der Waals correlation (BEEF-vdW) \cite{wellendorff2012density} provides a convenient and systematic way of performing realistic error estimates on the energies obtained from DFT calculations within the generalized gradient approximation (GGA).  The functional is built upon a combination of the reductionist and empiricist approaches. The exchange enhancement factor $F_x(s)$ in the GGA exchange energy density, $\epsilon_x^{GGA}(n,\nabla n)=\epsilon_x^{LDA}F_x(s[n,\nabla n)])$ is given by an expansion in terms of Legendre polynomials $B_m$ \cite{wellendorff2012density}
\begin{equation}
F_x^{GGA}(s)=\sum_m a_m B_m[t(s)]\;, 
t(s)= \frac{2s^2}{4+s^2},
\end{equation}
where $a_m$ are the expansion coefficients which are fitted using training datasets of quantities quantities representing chemistry, solid state physics, surface chemistry, and van der Waals interactions. Overfitting of properties from datasets is avoided by regularization of the GGA exchange expansion. Another parameter in the BEEF-vdW functional $\alpha_c$ arises in the correlation energy $E_c$ which has LDA, PBE and non-local contributions:
\begin{equation}
E_c=\alpha_c E^{LDA-c}+(1-\alpha_c) E^{PBE-c} + E^{nl-c}.
\end{equation}

To obtain uncertainty estimates on the DFT predicted energies, an ensemble of functionals around the optimum BEEF-vdW functional is used to calculate the energies non-self consistently. The ensemble of functionals is generated by creating a probability distribution for the model parameters $a_m$ and $\alpha_c$ such that the spread of the ensemble model predictions on the training data reproduces the errors obtained on using BEEF-vdW self-consistently.

\begin{figure}
\includegraphics[scale=0.45]{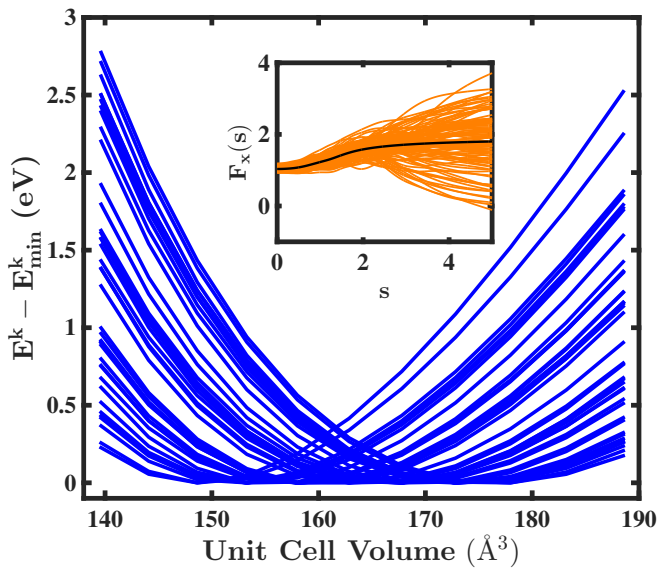}
\caption{\label{fig:FitEns}Ensemble of energy curves for Si obtained on applying a uniform strain in all three directions. The value of unit cell volume at the minima corresponds to the equilibrium volume for that ensemble. The inset shows an ensemble of 100 exchange enhancement factors $F_x(s)$ obtained on perturbing the values of coefficients $a_m$ of the Legendre polynomial $B_m$ in the exchange expansion. The optimum BEEF-vdW exchange enhancement factor is also shown (in dark) for comparison}
\end{figure}

For calculating the elastic constants, we fit the energies to the energy-strain relationship as discussed in section \ref{sec:propcalc}. Likewise, propagating uncertainty to the elastic constants would involve performing an ensemble of fits using the ensemble of energies generated at each point. The procedure is illustrated in Fig. \ref{fig:FitEns}.
The strain type is a uniform strain in all three directions that enables the computation of the unit cell volume and the bulk modulus. From the ensemble of energies generated at different values of the unit cell volume, we can bracket the minima for each such ensemble. This can be used to determine the unit cell volume (lattice constant) and bulk modulus for the corresponding density functional.

The procedure can be summarized as follows:
\begin{enumerate}
\item Choose the values $\zeta=\zeta_1,\zeta_2, \dots ,\zeta_n$ of the independent variable (strain parameter $x$ or unit cell volume $V$) to be used for fitting. Compute the transformation matrix $\rttensor{F}$ for each case.
\item Apply the homogeneous transformation $\rttensor{F}$ to the unit cell in each case. Relax the internal coordinates until the force is lower than the maximum allowed force, and calculate the energy.
\item Generate an ensemble of $m$ energies at each $\zeta$ using BEEF-vdW and perform the fitting using the relationship between energy and $\zeta$. For the $k^{th}$ ensemble, the fitting can be performed using the array of energies of that ensemble  $\rtvector{\mathfrak{E}}^k=[E^k(\zeta_1),E^k(\zeta_2),\cdots ,E^k(\zeta_n)]$.
\item Calculate the values of the fitting parameters $c^k_1, c^k_2, \cdots c^k_l$ for the $k^{th}$ ensemble using the $E-\zeta$ relationship and use them to calculate the elastic constants for the $k^{th}$ ensemble.
\item Repeat the process over all $m$ ensembles and generate the ensemble of elastic constants $C_{ij}$.
\end{enumerate}

It should be noted that the generation of ensembles for the energies and elastic constants is only a post-processing step consuming minimum time.  The proposed method has a distinct computational advantage over the approach of carrying out self-consistent energy calculations with several exchange-correlation functionals. The values of $m$ and $n$ need to be chosen carefully to get the correct uncertainty estimates. The value of $m$, the number of ensemble functionals used, should be chosen such that the uncertainty estimate is well-converged. The value of $n$ should be chosen such that the values of the fitting parameters are converged and the extreme values of $\zeta$ are able to provide a good fit for each of the ensembles, so that the minima for all ensembles falls within the extreme values of $\zeta$.

\section{Results and Discussion}
We begin by demonstrating our method using silicon as a material that has received considerable attention in experimental and DFT studies. We, then apply our method to a problem of current interest in the battery community.

\subsection{Test Case: Si}
\begin{figure}
\includegraphics[scale=0.42]{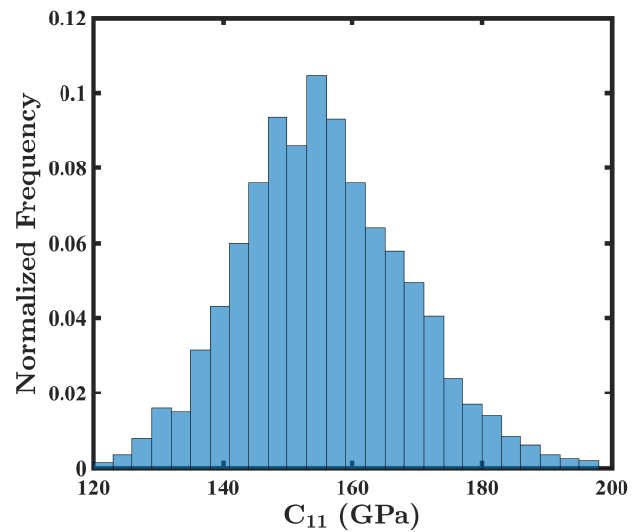}
\caption{\label{fig:SiC11}Distribution of values for the elastic constant $C_{11}$ of Si obtained using our method. The distribution has a standard deviation of 12.9 GPa and is a measure of the uncertainty in $C_{11}$.}
\end{figure}

In order to test whether our developed method indeed predicts the uncertainty associated with the elastic constants, we performed two sets of calculations for Si.  First, using our method, we calculate the elastic constants and the associated uncertainty.  Second, using a few different GGA exchange-correlation functionals, we calculate the elastic constants. The robustness of the uncertainty estimate is determined by its ability to bound the range of calculated elastic constants.  Using BEEF-vdW, we obtained an ensemble of values for each of the elastic constants of Si. The distribution of values obtained for the elastic constant $C_{11}$ using the developed method is shown in Fig. \ref{fig:SiC11}.

Table \ref{tab:Si} shows the values of unit cell volume and the elastic constants of Si calculated using BEEF-vdW and three GGA functionals: PBE, \cite{GGAPBE} PBEsol \cite{GGAPBEsol} and RPBE. \cite{GGARPBE} We used the standard deviation of the distribution of the property values as a measure of the uncertainty. From the table, it is clear that the uncertainty values obtained using the developed method accurately depict the variation in properties due to the choice of exchange-correlation functional.  We would also like to emphasize that the uncertainty obtained through the developed method is quite tight in all cases.

\begin{table}[b]%The best place to locate the table environment is directly after its first reference in text
\caption{\label{tab:Si}%
Elastic constants and equilibrium unit cell volume of Si using our method and different GGA exchange-correlation functionals. The uncertainty estimate used is the standard deviation associated with the elastic constant distribution. The values predicted by different GGA exchange-correlation functionals clearly lie within the uncertainty estimates.}
\begin{ruledtabular}
\begin{tabular}{lcccc}
\textrm{Property}&
\textrm{Present work}&
\textrm{PBE}&
\textrm{PBEsol}&
\textrm{RPBE}\\
\colrule
Volume ($\mathrm{\AA^3}$) & $164.3\pm6.7$ &164.2 & 160.9 & 167.1\\
$B$ (GPa) & $89.5\pm9.0$ & 88.7 & 93.8 & 94.6 \\
$C_{11}$ (GPa) & $155.1\pm12.9$ &  152.8 & 156.0 & 148.5\\
$C_{12}$ (GPa) & $56.7\pm7.8$ & 56.7 & 62.7 & 52.7\\
$C_{44}$ (GPa) & $77.1\pm4.0$ & 75.8 & 73.9 & 75.0\\

\end{tabular}
\end{ruledtabular}
\end{table}

\subsection{Application: Solid ion Conductors}
Next, we proceed to apply the developed method to the calculation of mechanical properties of solid ion conductors.  The mechanical properties of a solid ion conductor used in a Li-ion or Na-ion battery are important for its robust functioning and performance under the strains encountered during cycling. These strains typically arise due to volumetric expansion of the electrodes during intercalation. During such strains, the solid ion conductor should be able to maintain contact with the electrodes without substantial mechanical degradation. Another potential application of a solid ion conductor is to enable Li and Na metal anode by suppressing dendrites at its interface with the electrode. The suppression of dendrites has been linked to the shear modulus of the solid ion conductor. It has been found theoretically that a solid ion conductor with a shear modulus roughly twice that of Li at a Poisson's ratio of 0.33 can suppress Li dendrites. \cite{Monroe01022005}. Further, dendritic growth has also been shown to be affected by the Young's modulus of the solid ion conductor.\cite{Ferrese01012014} Stress generated at the solid ion conductor due to dendrite growth, higher than the yield strength of Li can result in suppression of Li dendrites through plastic deformation and flattening of the Li metal anode.

\begin{figure}
\includegraphics[scale=0.35]{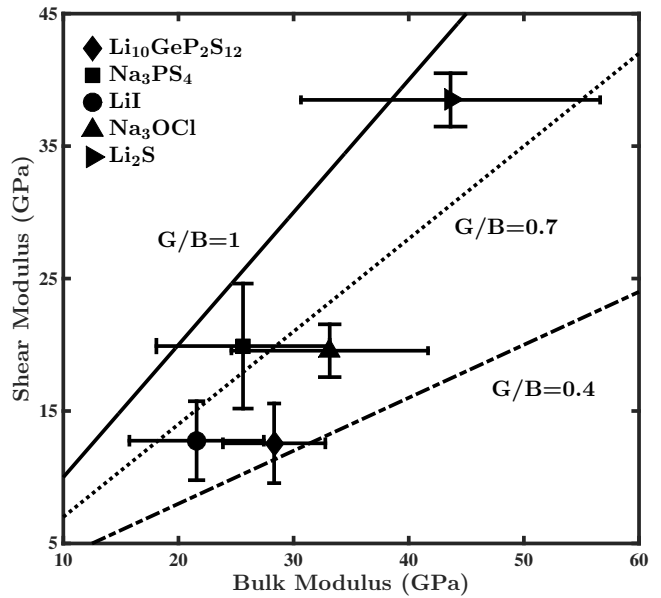}
\caption{\label{fig:Pughratio}A plot of shear modulus vs. bulk modulus with uncertainties for different solid ion conductors. The straight lines represent points with constant Pugh's modulus ratios.}
\end{figure}

We computed the elastic constants of solid ion conductors belonging to four different classes: thiophospahate, antiperovskite, glass and halide. The results of the calculations are tabulated in Table \ref{tab:solelec}.
\begin{table*}
\caption{\label{tab:solelec}Calculated elastic constants, bulk, shear, Young's moduli and Poisson's ratio for different classes of solid ion conductors. The uncertainty estimates used are the standard deviations associated with the distribution.}
\begin{ruledtabular}
\begin{tabular}{lccccccr}

Material & Volume ($\mathrm{\AA^3}$) & $C_{ij}$ (GPa) & $B$ (GPa) & $G$ (GPa) & $Y$ (GPa) & $\nu$
\\ \hline
\begin{tabular}{l}
$\mathrm{Li_{10}GeP_2S_{12}}$  \\ 
Thiophosphate \\
($\mathrm{Li_{10}MP_2S_{12}}$)
\end{tabular} & 

$979.8\pm43.2$ &

\begin{tabular}{cc}\\
$C_{11}:$&$46.5\pm5.6$\\
$C_{12}:$&$29.6\pm4.3$\\
$C_{13}:$&$13.6\pm 5.0$\\
$C_{33}:$&$49.5\pm 6.6$\\
$C_{44}:$ &$12.1\pm7.2$\\
$C_{66}:$ &$12.5\pm1.7$\\
 \end{tabular}  & $28.3\pm4.4$ & $12.6\pm3.0$ & $32.8\pm6.9$ & $0.31\pm0.05$\\
\begin{tabular}{l}
 $\mathrm{Na_3PS_4-I\rtvector{4}3m}$ \\
 Thiophosphate \\
 (Na-ion)
 \end{tabular}
 &
$350.9\pm21.2$ &
\begin{tabular}{cc}\\
$C_{11}:$&$50.2\pm9.1$\\
 $C_{12}:$&$13.3\pm7.1$\\
 $C_{44}:$ &$20.9\pm9.4$\\
 \end{tabular}& $25.6\pm7.5$  &$19.9\pm4.7$ & $47.4\pm9.3$ & $0.19\pm0.10$ \\
 
\begin{tabular}{l}
LiI \\
Halide
\end{tabular} &
$220.8\pm13.2$ & 

\begin{tabular}{cc}\\
$C_{11}:$&$33.1\pm13.7$\\
 $C_{12}:$&$15.8\pm6.3 $\\
 $C_{44}:$ &$16.5\pm1.7$\\
 \end{tabular} & $21.6\pm5.8$ & $12.8\pm3.0 $ & $32.0\pm7.2$ & $0.25\pm0.06$\\
 
 \begin{tabular}{l}
  $\mathrm{Na_3OCl}$ \\
  Antiperovskite
 \end{tabular}
 &
$95.6\pm5.7$ & 
\begin{tabular}{cc}\\
$C_{11}:$&$70.0\pm12.9$\\
 $C_{12}:$&$15.2\pm8.0 $\\
 $C_{44}:$ &$15.8\pm1.3$\\
 \end{tabular}
 
 & $33.1\pm8.5$ & $19.6\pm2.0$ & $49.0\pm5.5$ & $0.25\pm0.07$  \\

 \begin{tabular}{l}
 $\mathrm{Li_2S}$ \\
 Glass
 \end{tabular}
 &
$187.7\pm12.3$ &
\begin{tabular}{cc}\\
$C_{11}:$&$90.8\pm14.4$\\
$C_{12}:$&$20.0\pm12.4 $\\
$C_{44}:$&$40.7\pm3.2$\\
 \end{tabular} & $43.6\pm13.0$ & $38.5\pm2.0$ & $89.2\pm8.2$ &  $0.16\pm0.08$ \\
\end{tabular}
\end{ruledtabular}
\end{table*}
Most of the property values predicted using BEEF-vdW are in reasonable agreement with previous DFT calculations on solid ion conductors whenever available. \cite{Deng01012016, wang2014elastic} In many cases, significant uncertainty exists in the elastic moduli due to the exchange-correlation functional. Further, we propagated these uncertainties to a property of interest in solid ion conductors for batteries.  The shear modulus is plotted against the bulk modulus in Figure \ref{fig:Pughratio} along with constant Pugh's modulus ratio ($G/B$) lines. The Pugh's modulus ratio is a measure of the brittleness of the material. \cite{Pugh1954} Among the solid ion conductors we studied, $\mathrm{Li_2S}$ has the highest Pugh's modulus ratio and lies in the brittle regime. However, the uncertainty in the Pugh's modulus ratio predicted from DFT deserves attention due to the nature of the bounds which, in most cases cross the critical Pugh's modulus ratio of 0.571. \cite{Pugh1954}.

Another property of interest for a solid ion conductor is its ability to resist formation of dendrites or smoothen the roghness at the anode due to uneven deposition of Li. We calculated the Poisson's ratio and shear modulus of potential materials for solid ion conductors (Table \ref{tab:solelec}) which can be used to determine the dendrite suppressing ability of the material.\cite{Monroe01022005} We believe that the developed method of quantifying uncertainty will play a crucial role in large scale screening studies for desired mechanical properties. In particular, this method will be critical for the identification of a solid state Li-ion or Na-ion conductor that can mechanically suppress dendrites.

\section{Summary and Conclusions}
We have developed a method for obtaining uncertainty estimates on the mechanical properties predicted by first-principles calculations. We have demonstrated that the uncertainty estimates obtained through this method bound the property values calculated using several GGA functionals for Si.  This allows us to isolate the error in DFT calculations due to the choice of exchange-correlation functional. We applied this method to compute the mechanical properties of different classes of solid ion conductors with uncertainty. The advantage of our method is that different properties of practical interest can be predicted with the confidence that their value will lie within the uncertainty bounds, thereby avoiding multiple self-consistent calculations using several exchange-correlation functionals. We believe that the uncertainty estimation capability will dramatically increase the success of large-scale screening studies for desired mechanical properties.

\begin{acknowledgments}
The authors wish to acknowledge helpful discussions with Alan McGaughey.  Acknowledgment is made to the Donors of the American Chemical Society Petroleum Research Fund for partial support of this research.
\end{acknowledgments}

\bibliography{apssamp}

\end{document}